\documentstyle[fullpage,fleqn]{article}
\font\fonta=cmr12 scaled\magstep2

\font\fontc=cmr12 scaled\magstep1

\def\PL #1 #2 #3 {Phys. Lett.~{\bf#1} (#2) #3}
\def\NP #1 #2 #3 {Nucl. Phys.~{\bf#1} (#2) #3}
\def\ZP #1 #2 #3 {Z.~Phys.~{\bf#1} (#2) #3}
\def\PR #1 #2 #3 {Phys. Rev.~{\bf#1} (#2) #3}
\def\PRD #1 #2 #3 {Phys. Rev.~D {\bf#1} (#2) #3}
\def\PP #1 #2 #3 {Phys. Rep.~{\bf#1} (#2) #3}
\def\PRL #1 #2 #3 {Phys. Rev.~Lett.~{\bf#1} (#2) #3}
\def\RMP #1 #2 #3 {Rev. Mod. Phys.~{\bf#1} (#2) #3}

\def\ibid {{\it ibid}.}
\begin{document}
{
\begin{flushright}{TMU-NT930301\\March 1993}
\end{flushright}
}
\vglue 0.3cm
\large
\baselineskip=1cm
\begin{center}{\fonta Pion Structure Function \\
               in the Nambu and Jona-Lasinio model} \\
\vglue1.5cm
{\fontc 
Takayuki Shigetani\footnote{\large{e-mail 
address : shige@atlas.phys.metro-u.ac.jp}}
,  Katsuhiko Suzuki 
and  Hiroshi Toki\footnote{\large{Also at RIKEN, Wako, Saitama, 351, 
                             Japan}}\\
        {\em Department of Physics, Tokyo Metropolitan University}\\
        {\em Hachiohji,Tokyo 192, Japan}}
\vglue1.5cm
(\PL B308 1993 383 )
\vglue 3cm
\end{center}
\vglue 0.5cm
\baselineskip=0.8cm
\noindent
Abstract :  The pion structure function is studied in the Nambu and 
Jona-Lasinio (NJL) model.  
We calculate the forward scattering amplitude of a virtual photon 
from a pion target in the Bjorken limit, and obtain valence quark 
distributions of the pion at the low energy hadronic scale, where the 
NJL model is supposed to work.  
The calculated distribution functions are evolved to the experimental 
momentum scale using the Altarelli-Parisi equation.  
The resulting distributions are in a reasonable agreement with 
experiment.  
We calculate also the kaon structure function and compare the 
ratio of kaon to pion valence u-quark distributions with experiment.
\newpage
\baselineskip=0.8cm
Recent experimental observations of deep inelastic scattering (DIS) 
provide us with a large amount of data about the structure of hadrons.  
These results \cite{EMC,NMC} indicate that our understanding of hadron 
structure is still incomplete.  
Quantum chromodynamics (QCD) is believed to be the theory of the 
strong interaction, and perturbative QCD is consistent to the 
measured \(Q^2\) dependence of the structure function \cite{Reya}.  
However, QCD is not able to predict directly the structure function 
itself, since non-perturbative effects of confinement and dynamical 
symmetry breaking are not yet understood in the context of QCD.  
On the other hand, phenomenological quark models, which possess some 
non-perturbative features of QCD, give remarkable predictions of static 
hadron properties such as masses and magnetic moments.  
Hence, it is of our great interest to clarify a connection 
between the DIS phenomena and the low energy non-perturbative features 
in terms of such models as effective theories of QCD.  
Such a connection would enable us to use the DIS data to constrain 
the models of low energy QCD.

Recently, the nucleon structure functions are discussed in terms of the 
MIT bag mode \cite{4,5} as well as other models \cite{6}.  
Those results are based on the assumption that the structure functions 
at low energy hadronic scale \(Q^2 = Q_0^2\) (unknown) are obtained by 
calculating the twist 2 matrix elements within the effective models, 
since the twist 2 operators are dominant in the Bjorken limit and higher 
twist terms vanish at high enough \(Q^2\) \cite{4}.  
Once one obtains them at the hadronic scale, these parton distributions 
are evolved to the experimental scale by using the perturbative QCD, 
and the comparison with experiment can be made.

In this letter, we concentrate on the pion structure function.  
The pion is believed to be the Goldstone boson due to the spontaneous 
breakdown of the chiral symmetry (SB\(\chi\)S).  
Thus, the information of the pion structure function gives us deeper 
understandings of SB\(\chi\)S.  
The experimental data of the pion structure function are extracted from 
\(\pi\)N scattering using the Drell-Yan process \cite{7}.  
The valence, sea quark, and gluon distribution functions are obtained 
from a recent re-analysis of these data \cite{8}.

We use the Nambu and Jona-Lasinio (NJL) model \cite{9}, where the chiral 
invariance is the main ingredient.  
The NJL model describes the pion as a collective \(q\bar{q}\) excitation of 
the non-perturbative QCD vacuum \cite{10}.  
Recently, the generalized SU(3)\(_f\)  NJL model with the U(1)\(_A\) 
anomalous term is demonstrated to reproduce the meson properties 
successfully, in spite of the lack of confinement \cite{10,11,12}.  
This model is also applied to the chiral phase transition at finite 
temperature and density \cite{13}.  
All these results indicate that the NJL model possesses the essential 
features of QCD dynamics.

The SU(3)\(_f\)  NJL lagrangian is given by \cite{10,11,12}
%
%
\vglue 0.3cm
\begin{eqnarray}
 {\cal L}_{NJL}&=&\bar{\psi}(i\gamma^{\mu}\partial_{\mu}-m)\psi\nonumber\\
          & &+G_{S}[(\bar{\psi}\lambda_{i}\psi)^2
                  +(\bar{\psi}i\gamma_{5}\lambda_{i}\psi)^2]
          -G_{V}[(\bar{\psi}\gamma_{\mu}\lambda_{i}\psi)^2
                  +(\bar{\psi}\gamma_{\mu}\gamma_{5}\lambda_{i}\psi)^2]\; ,
\label{qcdlag}
\end{eqnarray}
\vglue 0.3cm
\noindent
where \(\psi\) denotes the quark field with current mass \(m\), 
\(\lambda_{i}\) are SU(3) flavor matrices, and \(G_S\), \(G_V\) 
are the coupling constants. 
In this model, the quarks acquire the constituent masses 
dynamically due to SB\(\chi\)S.  
The meson masses are obtained by solving the Bethe-Salpeter equation.  
This model describes the SU(3)\(_f\) meson properties very well with 
the parameters fixed by the pion and kaon properties \cite{10,11,12}.  
Using this lagrangian, we will calculate the parton distribution 
of pion with no free parameter.

Unfortunately, the NJL model is a non-renormalizable theory, and this 
model requires the finite momentum cutoff \(\Lambda\sim1\) GeV, 
which is identified with a scale of SB\(\chi\)S. 
One may understand the physical meaning of the cutoff as an approximate 
realization of "asymptotic freedom" in the NJL model.  
This means that the interaction between two quarks with the relative 
momentum larger than \(\Lambda\) is turned off, and two particles are free 
in such a high momentum scale.  
Hence, the structure functions obtained in the NJL model exhibit the 
Bjorken scaling in the deep inelastic limit \cite{11}.  
If we use the cutoff in a Lorentz invariant manner, we can calculate 
the quark distributions as a function of the Bjorken \(x\).

As mentioned above, we would like to calculate the structure function 
in the NJL model at a boundary to the QCD perturbation.  
The hadronic tensor \(W_{\mu\nu}\) is written by the structure functions 
\(F_1\) and \(F_2\) in the scaling limit;
%
%
\vglue 0.3cm
\begin{eqnarray}
W_{\mu\nu}=&-&(g_{\mu\nu}-\frac{q_{\mu}q_{\nu}}{q^2})F_{1}(x)\nonumber\\
           &+&\frac{1}{m_{\pi}\nu}(p_{\mu}-\frac{p\cdot q}{q^2}q_{\mu})
                           (p_{\nu}-\frac{p\cdot q}{q^2}q_{\nu})F_{2}(x)\; ,
\label{hadten}
\end{eqnarray}
\vglue 0.3cm
\noindent
where
%
%
\vglue 0.2cm
\begin{eqnarray*}
F_{2}(x)=x\sum_{i}e_{i}^{2}[q_{i}(x)+\bar{q}_{i}(x)]\; ,\;\;
F_{1}(x)=\frac{F_{2}(x)}{2x}\; .
\end{eqnarray*}
\vglue 0.2cm
\noindent
\(q_{i}(x)\) and \(\bar{q}_{i}(x)\) are the momentum distributions of 
the i-flavor quark and antiquark. 
The hadronic tensor is related to the forward scattering amplitude 
\(T_{\mu\nu}\) through the optical theorem \cite{Reya}. 
%
%
\vglue 0.3cm
\begin{eqnarray}
W_{\mu\nu}=\frac{1}{2\pi} \mbox{Im} T_{\mu\nu}\; .
\label{optical}
\end{eqnarray}
\vglue 0.3cm
\noindent
Thus, we evaluate \(T_{\mu\nu}\) in the NJL model to get 
the structure functions. 
The forward scattering amplitude of a virtual photon with \(q\) from 
a pion is defined by
%
%
\vglue 0.3cm
\begin{eqnarray}
T_{\mu\nu}=i\int d^{4}\xi e^{iq\xi}
\langle p\vert T[J_{\mu}(\xi)J_{\nu}(0)]\vert p\rangle_{C}\; ,
\label{forward}
\end{eqnarray}
\vglue 0.3cm
\noindent
where \(\vert p\rangle\) is a pion state of momentum \(p_{\mu}\) , 
and \(q_{\mu}\) the momentum delivered 
by the current.  The current is \(J_{\mu}=\bar{\psi}\gamma_{\mu}Q\psi\) 
with \(Q\) being the charge 
operator.  We compute the "handbag diagrams" for \(T_{\mu\nu}\), 
which is illustrated in fig.1a and 1b.  
The matrix element (\ref{forward}) in the NJL model is 
given by
%
%
\vglue 0.3cm
\begin{eqnarray}
T_{\mu\nu}=i\int \frac{d^{4}k}{(2\pi)^4}
\mbox{Tr}[\gamma_{\mu}Q\frac{1}{k\kern -2mm /}\gamma_{\nu}QT_{-}]+T_{+}
 \;\mbox{term}\; ,
\label{forward2}
\end{eqnarray}
\vglue 0.3cm
\noindent
where 
%
\vglue 0.2cm
\begin{eqnarray}
T_{-}=S_{F}(k-q)g_{\pi qq}\tau_{+}i\gamma_{5}S_{F}(k-q-p)
g_{\pi qq}\tau_{-}i\gamma_{5}S_{F}(k-q)\; .
\label{tminus}
\end{eqnarray}
\vglue 0.2cm
\noindent
Here, \(g_{\pi qq}\) is the coupling constant among two quarks and 
a pion obtained in the NJL model \cite{10,11,12}, 
and \(\tau_{\pm}=(\tau_{1}\pm i\tau_{2})/\sqrt{2}\).  
\(T_{-}\) and \(T_{+}\) denote scattering of a pion from quark and 
antiquark, respectively, where \(T_{+}\) has a similar expression as 
(\ref{tminus}).  
Note that the quark propagator contained in \(T_{-}\) has the 
dynamically generated mass obtained by solving the gap equation 
\cite{9,10}.  
Thus, \(S_{F}(k)\) is written as
%
%
\vglue 0.3cm
\begin{eqnarray*}
S_{F}(k)=\frac{1}{k\kern -2mm / -M}\; ,
\end{eqnarray*}
\vglue 0.3cm
\noindent
where \(M\) is the constituent quark mass.

The structure function obtained by the calculation of the "handbag 
diagram", shown in figs.1a and 1b, coincides with the leading twist 
contribution of the operator product expansion in the Bjorken limit 
\cite{14}.  
Therefore, these diagrams are enough for our purpose, because the 
structure functions are dominated by the twist 2 contributions 
in the DIS limit.  
Higher twist terms may be negligible at sufficiently high \(Q^2\) 
\cite{4}.

Following the work of Landshoff {\it et al.} \cite{15}, we will carry 
out the integration of (\ref{forward2}) in the Bjorken limit; 
%
%
\vglue 0.3cm
\begin{eqnarray}
Q^{2}=-q^{2}\to\infty\; ,\; \; \nu=\frac{p\cdot q}{m_{\pi}}\to\infty\; ,\; \;
x=\frac{Q^{2}}{2m_{\pi}\nu}\,:\,\mbox{fixed}\;.\nonumber
\end{eqnarray}
\vglue 0.3cm
\noindent
Here \(x\) is the so-called Bjorken \(x\), and \(m_{\pi}\) the pion mass.  
We introduce the integral variables
%
%
\vglue 0.3cm
\begin{eqnarray}
k_{\mu}=zp_{\mu}+yq_{\mu}+\kappa_{\mu}\; ,
\label{sudakov}
\end{eqnarray}
\vglue 0.3cm
\noindent
where \(\kappa_{\mu}\) satisfies \(\kappa \cdot p=\kappa \cdot q=0\).  
Thus, \(\kappa_\mu\) is spacelike \((\kappa^{2}< 0)\) and is effectively 
two dimensional.  
Calculating the traces and neglecting irrelevant terms in the Bjorken 
limit \cite{15}, we obtain
%
%
\vglue 0.3cm
\begin{eqnarray}
T_{\mu\nu}&=&\frac{8}{9}\frac{i}{(2\pi)^4}N_{c}g_{\pi qq}^2
\int\!dzd{\bar{y}}d^{2}{\kappa}
       (-g_{\mu\nu}+\frac{2z}{m_{\pi}\nu}p_{\mu}p_{\nu}
       +\frac{1}{m_{\pi}\nu}(p_{\mu}q_{\nu}+p_{\nu}q_{\mu}))\nonumber\\
   & &\times [t_{1}(\mu^{2},s)+zt_{2}(\mu^{2},s)]\frac{1}{z-x}\;,
\label{int1}
\end{eqnarray}
\vglue 0.3cm
\noindent
where
%
%
\vglue 0.2cm
\begin{eqnarray*}
t_{1}(\mu^{2},s)&=&\frac{1}{(\mu^{2}-M^{2}+i\varepsilon)^{2}}
                        \frac{1}{s-M^{2}+i\varepsilon}(\mu^{2}-M^{2})\;,\\
t_{2}(\mu^{2},s)&=&\frac{1}{(\mu^{2}-M^{2}+i\varepsilon)^{2}}
                        \frac{1}{s-M^{2}+i\varepsilon}(s-M^{2}-p^{2})
\end{eqnarray*}
\vglue 0.2cm
\noindent
and
%
%
\vglue 0.2cm
\begin{eqnarray*}
& &\mu^{2}=(k-q)^{2}=z\bar{y}+{\kappa}^{2}\;,\\
& &s=(k-q-p)^{2}=(z-1)(\bar{y}-m_{\pi}^{2})+\kappa^{2}\; .
\end{eqnarray*}
\vglue 0.2cm
\noindent
We change the variable \(\bar{y}=2m_{\pi}\nu(y-1)+zp^{2}\) 
\cite{15}, and \(\mu^{2}\) and \(s\) are the invariant masses of 
the struck quark and spectator.  
Performing the z-integral in eq.(\ref{int1}), we find \cite{15,11}  
%
%
\vglue 0.3cm
\begin{eqnarray}
T_{\mu\nu}&=&-\frac{8}{9}\frac{1}{(2\pi)^3}N_{c}g_{\pi qq}^2
            [-g_{\mu\nu}+\frac{2z}{m_{\pi}\nu}p_{\mu}p_{\nu}
             +\frac{1}{m_{\pi}\nu}(p_{\mu}q_{\nu}+p_{\nu}q_{\mu})]\nonumber\\
& &\times \int\! d{\bar{y}}d^{2}{\kappa}[t_{1}(\mu^{2}\, ,\,s)+xt_{2}(\mu^{2}
\, ,\,s)]\; .
\label{int2}
\end{eqnarray}
\vglue 0.3cm
\noindent
Consider now the \(\bar{y}\)-integral in the complex \(\bar{y}\)-plane.  
The \(s\)-propagator has a pole at 
\(\bar{y}=(\kappa^{2}-M^{2})/(1-x)+m_{\pi}^{2}+i\varepsilon/(1-x)\), 
and the \(\mu\)-propagator has a double pole at 
\(\bar{y}=(M^{2}-\kappa^{2})/x-i\varepsilon/x\).  
For \(x > 1\) or \(x < 0\), all these singularities occur on the same 
side of the real \(\bar{y}\)-axis, 
and (\ref{int2}) gives a zero result \cite{15,14}.  
However, if \(0 \leq x \leq 1\), the integration over \(\bar{y}\) no longer 
vanishes.  
Integrating (\ref{int2}) by \(\bar{y}\) and taking the imaginary part, 
we can get the quark distribution by use of the optical theorem 
(\ref{optical}).  
We change the integral variable \(k\) to \(\mu^{2}\), we find the following 
expression for the quark distribution:
%
%
\vglue 0.3cm
\begin{eqnarray}
q(x)&{\propto}&-g_{\pi qq}^{2}\int\! d\mu^{2}[\frac{1}{\mu^{2}-M^{2}}
             -x\frac{p^{2}}{(\mu^{2}-M^{2})^{2}}]\nonumber\\
& &\hspace{3cm} \times \theta (m_{\pi}^{2}x(1-x)-xM^{2}-(1-x)\mu^{2})\; .
\label{valence}
\end{eqnarray}
\vglue 0.3cm
\noindent
Here, \(\theta\) is the usual step function which arises from 
the spacelike condition of \(\kappa\).  
It is easily shown that the structure functions obtained from 
(\ref{valence}) exhibit the Bjorken scaling \cite{11}.  
We identify (\ref{valence}) as the valence quark distribution of 
the pion \(q_{V}(x)\), with the normalization \(\int_{0}^{1}\! 
dxq_{V}(x)=1\).  
The integral (\ref{valence}) goes to infinity due to the 
non-renormalizability of the NJL model.  
Thus, we shall use the momentum cutoff.  
There exists an ambiguity in the introduction of the NJL cutoff for 
DIS phenomena \cite{16,16p}.  
We introduce the Euclidean momentum, 
\(t_{E}^{2}=-\mu^{2}+m_{\pi}^{2}x-xM^{2}/(1-x)\), 
and simply assume that the four momentum of the struck quark has 
the sharp Euclidean cutoff, \({\theta}({\Lambda}^{2}-t_{E}^{2})\), 
which is consistent to the NJL cutoff.  
We examine also the exponential cutoff used in ref.\cite{16} and other 
plausible forms for completeness.  
We obtain similar results with ours except for the 
behavior of $q(x)$ around $x\sim 1$ which depends on the cutoff scheme.

Integrating (\ref{valence}) by \(t_{E}(\mu^{2})\) with the cutoff, 
we obtain the valence distribution of pion at the low energy hadronic 
scale, which is shown in fig.2.  
Here, we use the NJL parameters fixed by the pseudoscalar meson 
properties 
\(\Lambda=900MeV\), \(G_{S}\Lambda^{2}=2G_{V}\Lambda^{2}=4.67\), 
\(m_{u}=m_{d}=5.5MeV\) and \(m_{s}=135MeV\) \cite{10,11,12}.  
With this parameter set, we obtain the constituent masses, 
\(M_{u}=M_{d}=350MeV\) and \(M_{s}=534MeV\).  
Our results on the quark distribution 
depend only slightly on the choice of the parameters within the reasonable 
parameter ranges, once these parameters are constrained by the pion mass 
and the decay constant.
Note that the resulting distribution shows a correct behavior around 
\(x\sim 1\) ; \(q_{V}(x)\to 0\) as \(x\to 1\) \cite{11}.  
It is the advantage of our use of the NJL model, which is Lorentz 
invariant, as compared with the MIT bag model where the translational 
invariance is broken \cite{5}.  
We also note that this low energy scale structure function has no 
physical meaning at this scale, since "real" structure function at the 
low energy scale receives non-negligible contributions from all twist 
operators.  
Our calculated result in fig.2 plays a role of only the boundary 
condition of the structure function for the \(Q^{2}\) evolution.

We take the low energy hadronic scale at \(Q_{0}^{2}=(0.5 GeV)^{2}\), 
which is used in ref.\cite{17}.  
At this scale, the running coupling constant is still small; 
\(\alpha_{s}(Q_{0}^{2})/\pi\sim0.3\).  
Indeed, the inclusion of the second order QCD corrections gives a small 
change for the \(Q^{2}\) evolution from our result within 10\% \cite{17}.  
We may understand intuitively the physical implication of this 
scale \(Q_{0}^{2}\) as compared to the valence quark core radius 
of the pion \(\langle r^{2}\rangle _{core}\), noted by Brown et al. 
\cite{18}.   
Their value is consistent to our low energy scale:
%
%
\vglue 0.3cm
\begin{eqnarray*}
\langle r^{2}\rangle _{core}\sim (0.35 \mbox{fm})^{2}\sim 1/(0.5GeV)^{2}\;.
\end{eqnarray*}
\vglue 0.3cm

We use the first order Altarelli-Parisi equation \cite{19,Reya} for 
the \(Q^{2}\) 
evolution of valence distributions with \(\Lambda_{QCD}=250MeV\).  
We show in fig.2 (solid curve) the result at \(Q^{2}=20GeV^{2}\), 
which is compared with the experimental curve (dash-dotted curve) 
extracted from the Drell-Yan process \cite{8}.  
We find a reasonable agreement with experiment.  
If we vary the low energy scale \(Q_{0}^{2}\) to a larger value 
\(Q_{0}^{2}=(0.75GeV)^{2}\), the change of our distribution is 
of order 20
distribution function at the experimental scale, 
it seems necessary to choose \(Q_{0}^{2}\) as small as 
\(Q_{0}^{2}=(0.5GeV)^{2}\).  
We compare the pion structure function \(F_{\pi}(x)\) with the experiment 
in fig.3.  
It shows a good agreements for \(x > 0.3\).  
In the low \(x\) region, the structure function is dominated by the 
sea quark distribution \cite{Reya}.  
We ought to calculate the sea quark contributions \cite{16,16p} with 
the next-to-leading order QCD perturbation, which is under consideration.

We also calculate the first two moments of the valence quark 
distributions, \(\langle x\rangle\) and \(\langle x^{2}\rangle\), 
which are defined as
%
%
\vglue 0.3cm
\begin{eqnarray*}
\langle x\rangle =\int_{0}^{1}\!dxxq_{v}(x)\; \; ,\; \;
\langle x^{2}\rangle =\int_{0}^{1}\!dxx^{2}q_{v}(x)\;.
\end{eqnarray*}
\vglue 0.3cm
\noindent
The resulting values are shown in table 1 with the experimental values 
\cite{8} and the results of the lattice QCD calculations \cite{20}.  
Our results reproduce the data very well.

Finally, we discuss the kaon structure function.  
We can extract the effects of the SU(3) flavor symmetry breaking from 
quark distributions of the kaon.  
The valence quark distributions are obtained in a similar manner 
using the NJL model.   
The u- and s-quark distributions of the kaon at the low momentum scale 
and those at \(Q^{2}=20GeV^{2}\) are shown in fig.4.  
Our results indicate that the heavy s-quark carries a larger fraction of 
kaon momentum than the light u(d) quark, as expected.  
The resulting distributions show a similar behavior with the previous 
results obtained by the analysis of the Regge trajectories \cite{21}.  
We also show in fig.5 the ratio of kaon to pion valence u-quark 
distributions \(u_{K}/u_{\pi}\) at \(Q^{2}=20GeV^{2}\) with 
experimental data \cite{22}.  
This ratio is sensitive to the mass difference of the constituent quark 
masses.  
In fact, our calculations give a simple expression of this ratio around 
\(x\sim 1\) at the low momentum scale as
%
%
\vglue 0.3cm
\begin{eqnarray*}
u_{K}/u_{\pi}\sim (M_{u}/M_{s})^{2}\sim 0.5\; .
\end{eqnarray*}
\vglue 0.3cm
\noindent
This value is quite reasonable in view of the experimental findings.

In conclusion, we have studied the pion structure function at the DIS 
scale using the NJL model as a low energy effective theory of QCD.  
The resulting distribution at the low momentum scale is evolved to the 
experimental scale, at which it compares with the DIS data very well.  
We have also studied the kaon structure function.  
Reflecting the SU(3) flavor symmetry breaking due to the bare current 
mass difference, the \(x\) dependence of the ratio of kaon to pion valence 
u-quark distributions is found consistent with experiment.  
Our results indicate that the low energy effective quark model is able 
to describe the DIS phenomena.  
We shall point out that the sea quark degrees of freedom can be 
naturally included in the framework of the NJL model, 
which is needed to compare with the pion structure function directly.  
It is possible to calculate the sea quark distributions in the NJL 
model as higher loop corrections \cite{16,16p}.

We are grateful to T. Hatsuda and T. Kobayashi for illuminating 
discussions on the deep inelastic phenomena.
\newpage
\noindent
{\bf Table Caption\/}
\vglue 1cm
\begin{description}
\item[Table 1] : The first two moments of the pion distribution at 
\(Q^{2}=49 GeV^{2}\).  
The theoretical calculations of the NJL model are shown in the 
second row, and compared with the experimental values \cite{8} (third 
row) and the lattice QCD results \cite{20} (last row).  
\end{description}
\vglue 2cm
\begin{center}
{\bf Table 1}\\
\vglue 1cm
\begin{tabular}{c c c }\hline
{ }&{\(2\langle x \rangle\)}&{\(2\langle x^{2} \rangle\)}\\ \hline
Theory & 0.41 & 0.16 \\
Experiment & 0.40\(\pm\)0.02 & 0.16\(\pm\)0.01 \\ 
Lattice & 0.46\(\pm\)0.07 & 0.18\(\pm\)0.05 \\ \hline
\end{tabular}
\end{center}
\newpage
\noindent
{\bf Figure Captions}
\vglue 1cm
\begin{description}
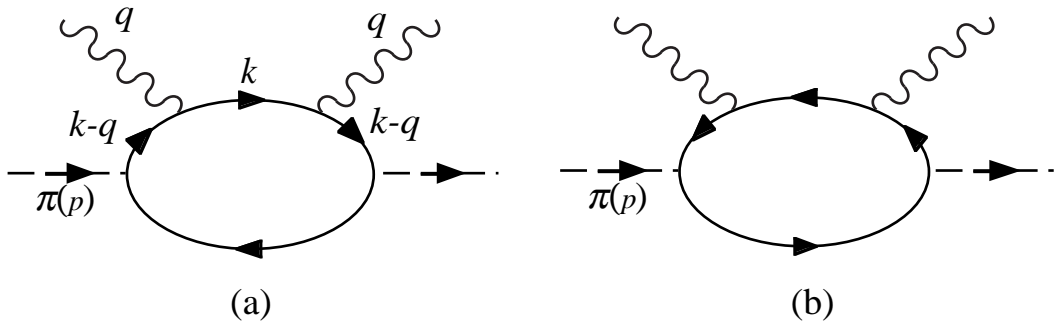
\item[Fig. 1]: The forward Compton scattering amplitude 
("handbag diagram") of the pion.  
The solid line represents the quark.  
The pion and the virtual photon are depicted 
by the dashed and wavy lines.  For 
details and notation, see text.
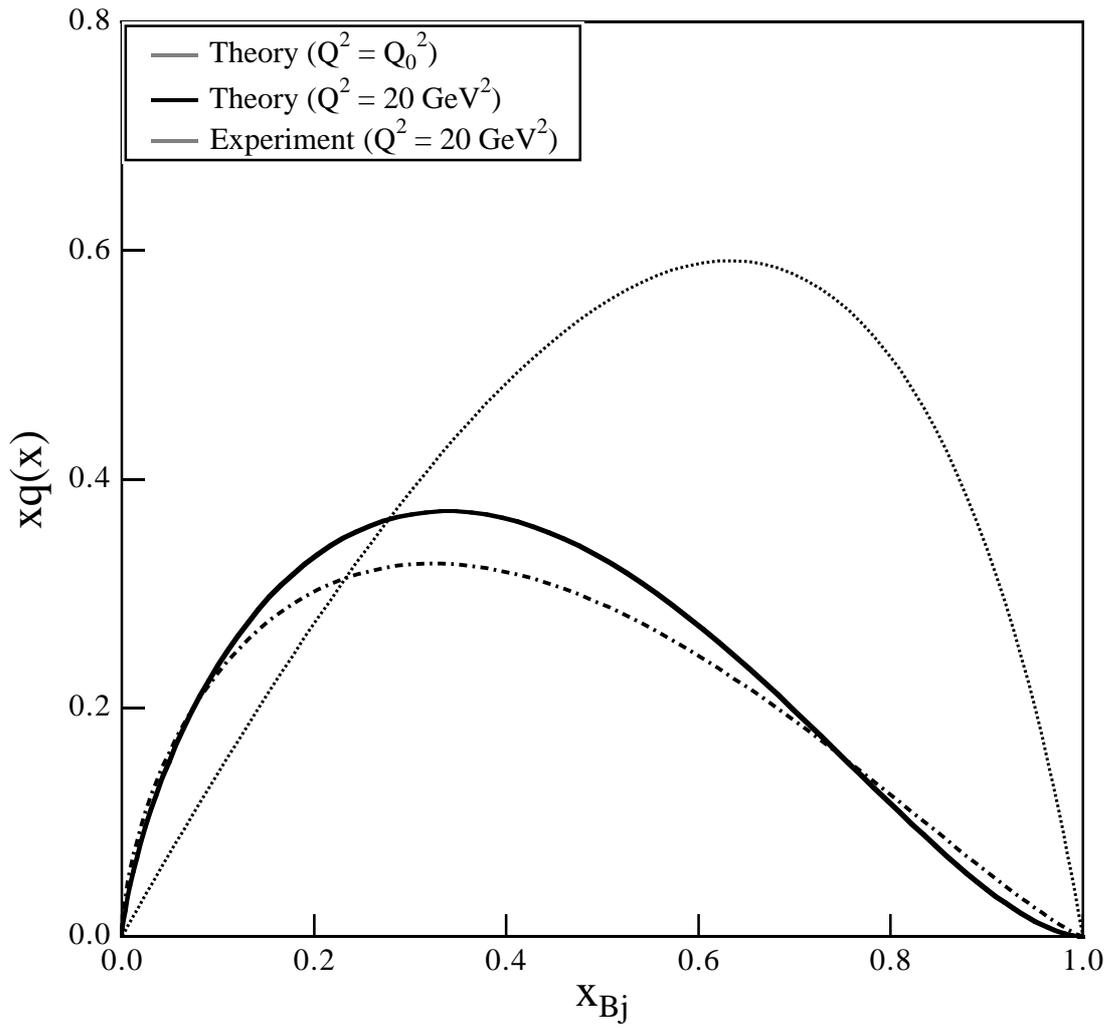
\item[Fig. 2]: The valence quark distributions of the pion 
at the low energy scale \(Q^{2}=Q_{0}^{2}\) (dotted curve) 
and at \(Q^{2}=20GeV^{2}\) (solid curve) as a function of the Bjorken $x$.  
The experimental fit \cite{8} is depicted by the dash-dotted curve.
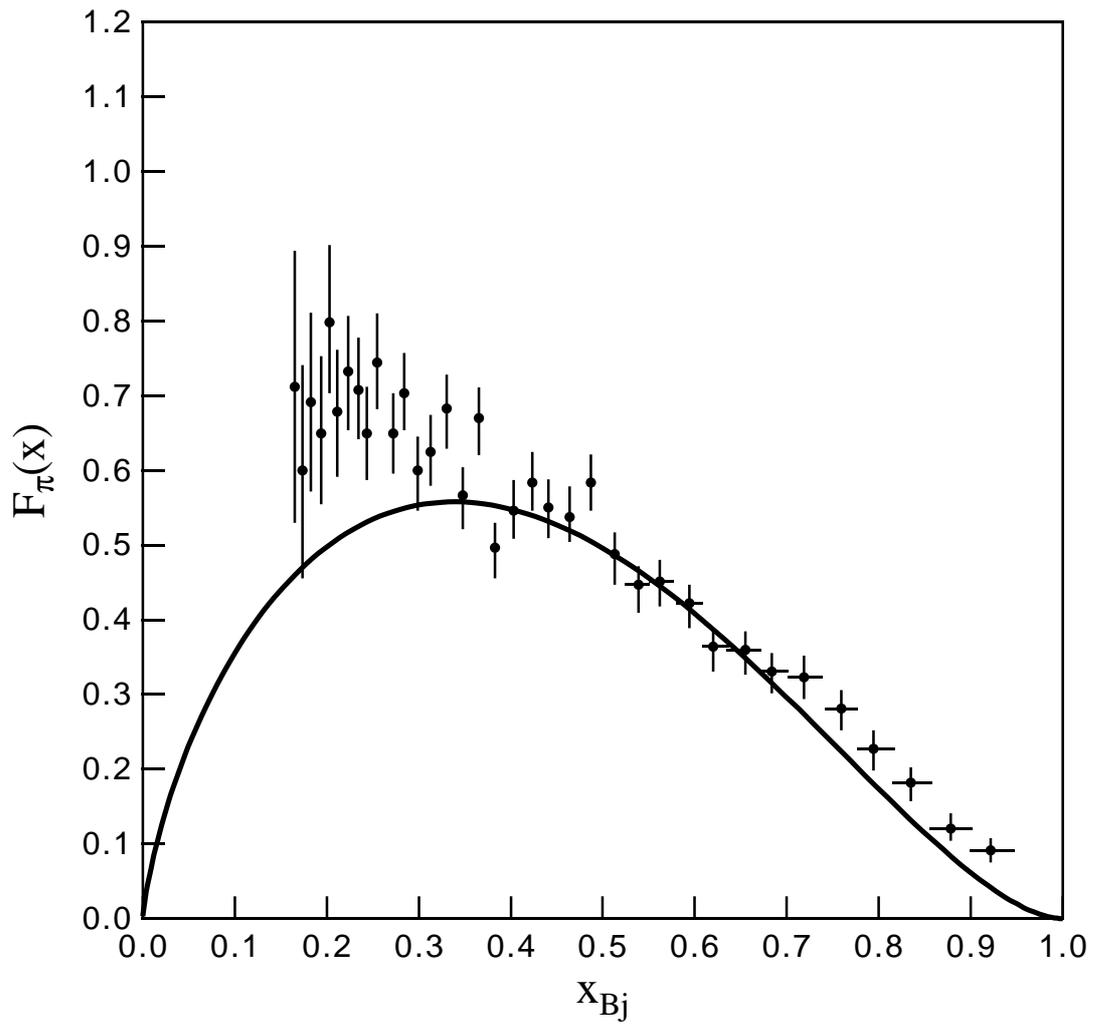
\item[Fig. 3]: The pion structure function at \(Q^{2}=20GeV^{2}\).  
The experimental data 
are taken from the NA3 experiment \cite{7}.  The theoretical prediction 
of the 
NJL model is depicted by the solid curve.  Here, we use 
\(F_{\pi}(x)=Kxq_{V}(x)\) with the sea quark distributions being neglected, 
where we take the \(K\)-factor, \(K\)=1.5 \cite{8}.
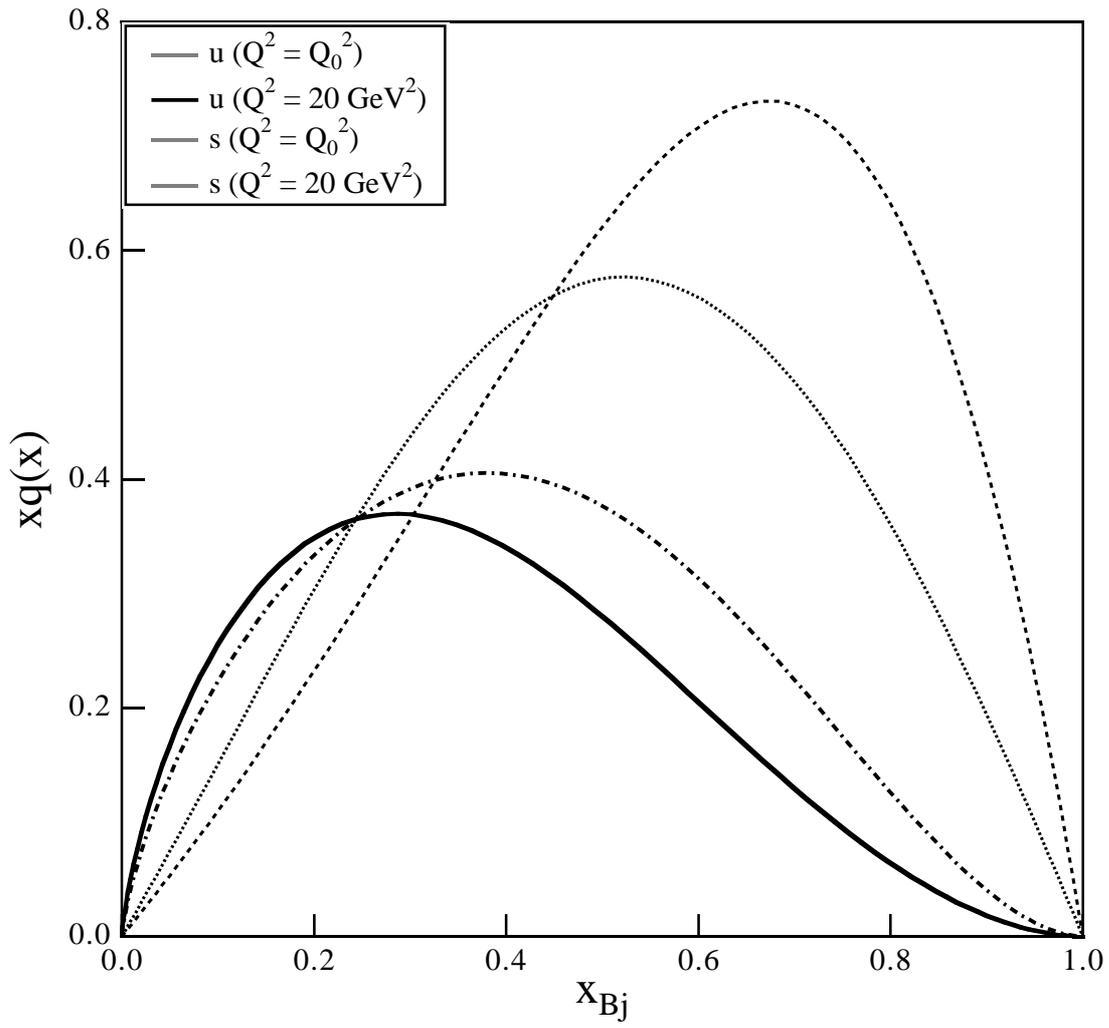
\item[Fig. 4]: The valence quark distributions of the kaon at low energy 
scale and at \(Q^{2}=20GeV^{2}\) in the NJL model.  
The u- and s-quark distribution at 
\(Q_{0}^{2}\) are depicted by the dotted and dashed curves. The solid and 
dashed-dotted curves represent the u- and s-quark valence distributions 
at \(Q^{2}=20GeV^{2}\).
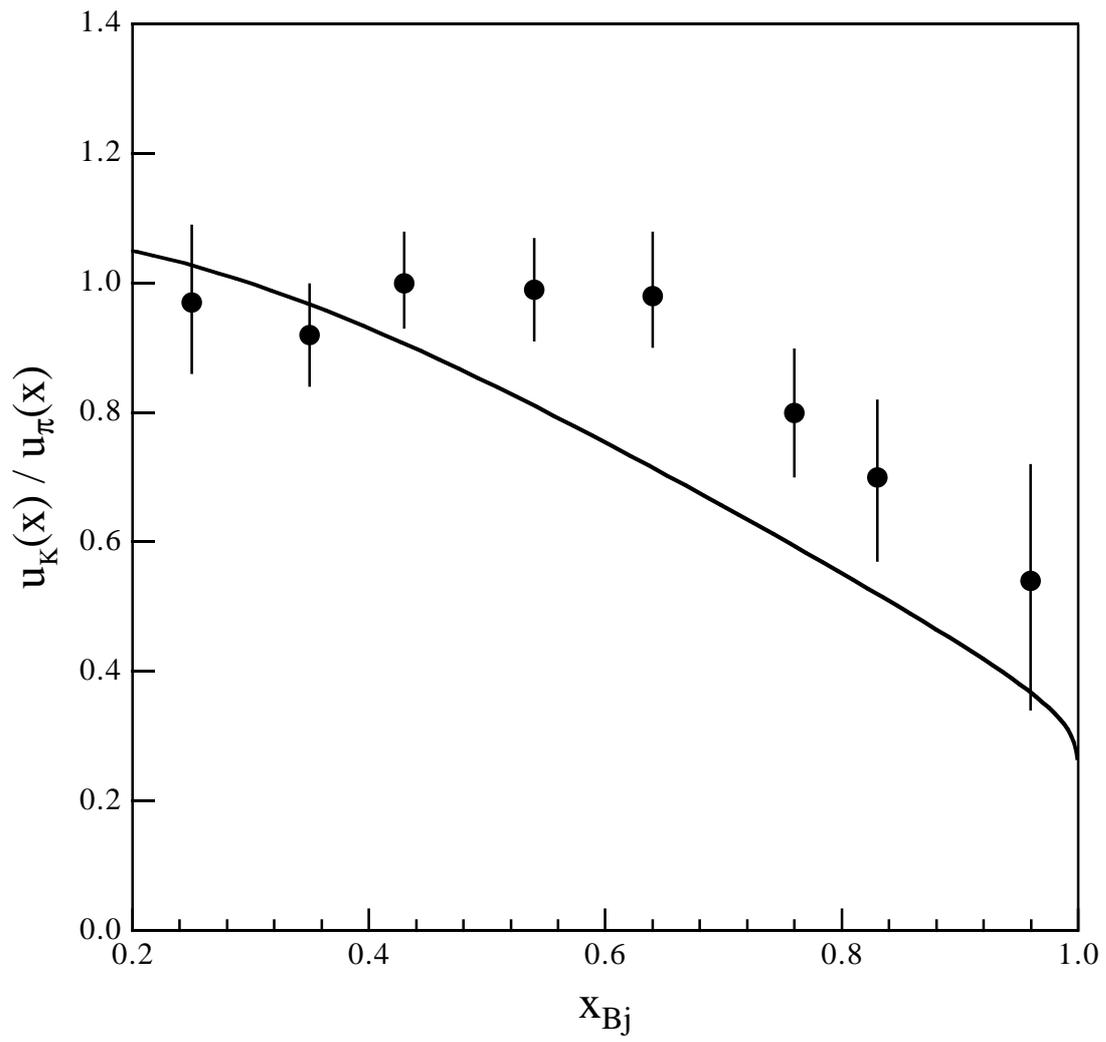
\item[Fig.5]: The ratio of kaon to pion valence u-quark distributions\\
 \(u_{K}/u_{\pi}\) at \(Q^{2}=20GeV^{2}\) scale.  
The theoretical result is depicted by the solid curve.  
The experimental values with error bars are taken from ref.\cite{22}.
\end{description}
\end{document}